\documentclass[12pt]{article}
\usepackage{varwidth}
\usepackage{cite}
\usepackage{physics}
\usepackage{url}
\usepackage{empheq}
\usepackage{wasysym}
\usepackage{marvosym}
\usepackage{halloweenmath}
\usepackage{staves}
\usepackage{amsthm,amsmath,latexsym,amssymb,amsfonts,amsbsy}
\usepackage{graphicx,lscape,fancyhdr,xcolor,array,stmaryrd,euscript,wrapfig}
\usepackage{hyperref}
\usepackage{tikz}
\usetikzlibrary{decorations.pathmorphing}
\newcommand{\ext}{{\rm d}}

\newcommand{\R}{\mathbb R}
\newcommand{\Z}{\mathbb Z}
\newcommand{\T}{{\rm Tr\,}}

\definecolor{darkmagenta}{rgb}{0.55, 0.0, 0.55}
\definecolor{darkblue}{rgb}{0.0, 0.0, 0.55}
\definecolor{darkred}{rgb}{0.7, 0.0, 0.3}
\hypersetup{
	unicode,	
	colorlinks,
	citecolor=darkred,
	linkcolor=darkblue, 
	urlcolor=darkred,
	bookmarksopen=true,
	bookmarksnumbered
	}

 \csname
@addtoreset\endcsname{equation}{section}
\begin{document}
\begin{titlepage}
\setcounter{page}{1}
\begin{center}
\hfill
\vskip 0.1cm
{\LARGE Extended AdS spacetime without boundaries \\[5pt]
and entanglement without holography}
\vskip 30pt
{\sc  Cesar Arias}
\vskip 5pt
{\it
Departamento de Matem\'atica, Pontificia Universidad Cat\'olica de Chile\\[-3pt]
Avenida Vicuña Mackenna 4860, Santiago, Chile
}
\vskip 5pt
\href{mailto:cesar.arias@uc.cl}{\texttt{cesar.arias@uc.cl}}
\vskip 40pt
{\bf Abstract}
\end{center}
\vskip -10pt
We glue together two copies of pure AdS spacetime along their conformal boundaries creating a manifold without boundaries.  
The resulting space, which in dimension $d+2$ we denote by $AdS^{d+2}_\pm$, has the topology of $S^2\times \Sigma^d$, where $\Sigma^d$ is a $d$-manifold without boundary.  Acting with~$\mathbb Z_n$ on the $S^2$ factor amounts to coupling a pair of membranes at the north and south poles of the 2-sphere. 
Moreover, extending the domain of the 2-sphere polar coordinate from $[0, \pi]$ to the interval $[0, (2N-1)\pi]$, where $N>1$,  enables the coupling of one stack of $N$ coincident membranes at each pole of the 2-sphere ($2N$ membranes in total). 
Assuming the existence of a quantum gravity theory on the glued spacetime, we compute the classical approximation of the entanglement entropy across an entangling surface consisting of the two antipodal stacks of membranes. 
We find that the resulting entropy exhibits a boundary cutoff divergence that can 
be canceled by taking the limit of an infinite number of membranes. 
This large-$N$ cancellation---possible only in the doubled, extended geometry without boundaries---yields a finite, universal quarter-area law.
The calculation does not require details of the quantum theory other than its infrared limit, which we assume to be Einstein gravity. 
\end{titlepage}
{\small\tableofcontents }       

\section{Introduction}

One of the most salient and universal features that arises from the interplay between quantum theory and gravity is the appearance of area laws for entropy. 
A prime example is the Bekenstein--Hawking formula~\cite{Bekenstein:1973ur, Hawking:1974sw} for the entropy of a black hole, whose origins and implications have driven an important part of the recent research in string theory and gravity, as it arguably provides a semiclassical window into quantum aspects of spacetime.

Originally derived in the context of black hole thermodynamics, the area law has since revealed itself in a broader array of settings. 
Within the AdS/CFT correspondence~\cite{Maldacena:1997re}, the Ryu--Takayanagi formula~\cite{Ryu:2006bv} generalizes the relation between entropy and area to quantum field theories with gravitational duals. The formula  equates the entanglement entropy of a region in the boundary dual theory to one-quarter of the area of a minimal surface in the bulk. 
This geometrization of entanglement has further fueled the idea that spacetime itself may be a manifestation of quantum entanglement~\cite{VanRaamsdonk:2010pw}, and the recurrence of the one-quarter prefactor suggests a universality that extends beyond the black hole thermal case.

Building on this framework, Lewkowycz and Maldacena~\cite{Lewkowycz:2013nqa} provided a derivation of the Ryu--Takayanagi formula from first principles within the gravitational path integral. 
By extending the replica trick~\cite{Calabrese:2004eu} to the bulk and analyzing replica-symmetric geometries, they demonstrated that the dominant saddle-point contribution to the path integral yields an area term from a codimension two surface fixed under the replica symmetry. 
In the semiclassical limit, this surface is precisely the extremal surface of the Ryu--Takayanagi prescription. This work grounded the relation between entropy and area in a more rigorous formulation of quantum gravity, reinforcing the ubiquity of the one-quarter of the area law in holographic settings.

Despite the central role of holography in crystallizing these ideas, many of the phenomena it captures appear to reflect more general principles. In particular, the appearance of area laws in vacuum entanglement entropy across a wide variety of quantum systems~\cite{Bombelli:1986rw, Srednicki:1993im, Casini:2009sr, Casini:2011kv, Eisert:2008ur, Gong:2017crn} suggests that the geometric nature of entanglement entropy may not require a dual boundary theory to be realized.  Similarly, black hole entropy, with its universal structure, emerges from semiclassical gravity without invoking any particular holographic argument.

This raises the following question: to what extent can the relation between entanglement entropy and the area of minimal surfaces be understood and derived in non-holographic settings?  In particular, can area laws of the Ryu--Takayanagi type be reproduced or reinterpreted intrinsically within the bulk spacetime without appealing to a dual boundary description?

The goal of this note is to argue that Ryu--Takayanagi-like formulas can arise from more direct gravitational and field-theoretic considerations in anti-de Sitter (AdS) spacetime, with no need of holography.

To this end, we construct a geometry consisting of two copies of AdS glued along their conformal boundaries. Unlike a single AdS copy, this doubled geometry allows for the coupling of an arbitrary number of membranes. These are minimal surfaces that appear as the fixed points of a bulk $\mathbb Z_n$ symmetry, and are organized into two antipodal stacks of $N$ coincident membranes each. We argue that, in a semiclassical approximation, the entanglement entropy between two spacetime subregions separated by the two stacks of membranes leads to a novel area law.

For a finite number $N$ of membranes, the entropy depends on a boundary cutoff and diverges as this cutoff is taken to zero. However, this divergence can be suppressed by taking the limit of an infinite number of membranes. In this limit, the resulting entanglement entropy becomes cutoff independent and obeys a Ryu--Takayanagi-like formula.

The article is organized as follows. In~\S\ref{Sec2} we construct our extended geometry by gluing along the boundary two copies of AdS space. 
In~\S\ref{Sec3} we show that the extended geometry naturally admits the coupling of an arbitrary number of membranes, organized into two antipodal stacks of $N$ coincident membranes each. 
In \S\ref{Sec4} we devise a spacetime subregion separated by the two stacks of membranes, and compute the associated entanglement entropy.
We conclude in \S\ref{Sec5} with some final comments and remarks.

\section{Extended AdS space without boundaries} 
\label{Sec2}
In this section we assemble two copies of AdS space by fully overlapping their conformal boundaries. This construction results in a negatively curved manifold without boundaries, which hereafter we refer to as extended AdS space without boundaries; in $d+2$ dimensions we denote this manifold by $AdS^{d+2}_\pm$.

We first focus on the lower dimensional cases and devise the two and three-dimensional extended spaces $AdS^2_\pm$ and $AdS^3_\pm$. We then use these as a seed to recursively define
%
\begin{equation}\label{adspm}
AdS^{d+2}_\pm = S^2 \times AdS^d_\pm ~,\qquad d\geq2~.
\end{equation}
As we will elaborate on further below, equation~\eqref{adspm} follows from the fact that a single copy of pure $AdS^{d+2}$ space can be realized as the product $D^2\times \Sigma^d$, where $D^2$ is a two-dimensional disk and $\Sigma^d$ a negatively curved $d$-manifold without boundary which we take to be $\Sigma^d=AdS_\pm^d$, and where thus the boundary of $AdS^{d+2}$ has the topology of $S^1\times\Sigma^d$. Hence, in this context, gluing two copies of AdS means taking the corresponding copies of the disks $D^2$ and gluing them along their boundary circle.

To begin with, let us recall some elementary facts. Anti-de Sitter spacetime of dimension $d+2$ can be defined as a hyperboloid embedded in flat ambient space $\R^{2, d+1}$. 
Taking the ambient space coordinates to be $X^A= (T^1, T^2, X^a)$, where $a=1, ..., d+1$, the ambient space metric
\begin{equation}\label{eta}
\eta = -(\ext T^1)^2 - (\ext T^2)^2  + \sum_{a=1}^{d+1} (\ext X^a)^2~.
\end{equation}  
The embedding $AdS^{d+2}\hookrightarrow \R^{2, d+1}$ is thus defined by means of the hypersurface equation
\begin{equation}\label{X2}
X^2= \eta_{AB} X^AX^B = -(T^1)^2 - (T^2)^2  + \sum_{a=1}^{d+1} (X^a)^2 = -L^2,
\end{equation}
where $L$ denotes the $AdS^{d+2}$ radius.  Different parametrization of the coordinates $X^A\in\R^{2, d+1}$ gives rise to different induced geometries on $AdS^{d+2}$, which are given by the pull-back of the flat metric~\eqref{eta} onto the hypersurface defined by~\eqref{X2}, namely
\begin{equation}\label{gads}
g_{AdS} = \eta \big\vert_{X^2=-L^2}~.
\end{equation}
The induced geometry~\eqref{gads} in turn induces the topology of AdS space.

\subsection{Extended $AdS^2$}
\label{Sec21}
Let us first consider the lowest dimensional case and construct the extended version of $AdS^2$. To this end, we take the coordinates $(T^1, T^2, X)\in\R^{2,1}$ to be parametrized as
\begin{equation}\label{par2}
T^{1} = L\sec\theta \cos(t/L)~,\qquad
T^{2}=L\sec\theta \sin(t/L) ~, \qquad
X  = L \tan\theta~,
\end{equation}
where $-\tfrac{\pi}{2} \leq \theta\leq \tfrac{\pi}{2}$ and $t\in\R$. It is direct to check that the choice above satisfies the hypersurface constraint~\eqref{X2}.  The corresponding induced metric on (a single copy of) $AdS^2$ is obtained by restricting the flat embedding space metric $\eta$ to the surface $X^2=-L^2$ by means of the the parametrization~\eqref{par2}. This gives
\begin{equation}\label{gads2}
g_{AdS} = \frac{L^2 \ext \theta^2 - \ext t^2}{\cos^2\theta}~.
\end{equation}     

The advantage of these coordinates is that we can now glue two copies of $AdS^2$ simply by extending the range of the polar coordinate from $[-\frac{\pi}{2}, \frac{\pi}{2}]$ to the interval $[-\frac{\pi}{2}, \frac{3\pi}{2}]$.  
Indeed, taking these two copies to be defined as 
\begin{align}\label{pmdef}
AdS^2_+&:=\bigg\{ (T^1, T^2, X)\in\R^{2,1} |-\frac{\pi}{2} \leq \theta \leq \frac{\pi}{2} \bigg\}~, \cr
AdS^2_-&:=\bigg\{(T^1, T^2, X)\in\R^{2,1} |\quad\,\frac{\pi}{2} \leq \theta \leq \frac{3\pi}{2} \bigg\}~,
\end{align}
then each of them has a boundary given by a pair of points (a 0-sphere) times $\R$; the boundary of  $AdS^2_+$ is defined by the pair $\{-\frac{\pi}{2}, \frac{\pi}{2}\}$, while the boundary of $AdS^2_-$ is determined by $\{\frac{\pi}{2}, \frac{3\pi}{2}\}$. Thus, the space
\begin{equation}
AdS^2_\pm = AdS^2_+ \cup AdS^2_-
\end{equation}
can be defined as the result of overlapping the two boundaries by pairwise identifying the four points that compose them, that is, identifying the pair of points located at $\theta=\frac{\pi}{2}$ and the pair of points located at $\theta=-\frac{\pi}{2}$ and $\theta=\frac{3\pi}{2}$ . 
Note that the range of the polar coordinate is chosen such that, in the resulting glued space, the north and south poles of the spatial circle are located at $\theta=0$ and $\pi$, respectively. 
 The construction is depicted in figure~\hyperlink{Fig:1}1. 
 %
\begin{center}
\begin{tikzpicture}
\draw[thick, darkblue] (0,0) arc[start angle=0, end angle=180, radius=1.5cm];
\fill[darkred] (-3,0) circle (2pt); 
\fill[darkred] (0,0) circle (2pt); 
\node at (-3,1.5) {$AdS^2_+$};
\draw[thick, darkblue] (-3,-0.7) arc[start angle=180, end angle=360, radius=1.5cm];
 \fill[darkred] (-3,-0.7) circle (2pt); 
 \fill[darkred] (0,-0.7) circle (2pt); 
\node at (0.2,-1.9) {$AdS^2_-$};
\draw [magenta,<->, very thick] (-3.2,-0.1) to [bend right] (-3.2,-0.6);
\draw [magenta,<->, very thick] (0.2,-0.1) to [bend left] (0.2,-0.6);
\node at (-0.3,0) {$\frac\pi2$};
\node at (-0.3,-0.7) {$\frac\pi2$};
\node at (-2.6,0) {-$\frac\pi2$};
\node at (-2.55,-0.7) {$\frac{3\pi}{2}$};
%
\node at (-4,-0.4) {\footnotesize{gluing}};
\node at (1,-0.4) {\footnotesize{gluing}};
\node at (5,1.6) {$AdS^2_\pm$};
\draw[thick, darkblue] (5,-0.3) circle (1.5);
 \small
\node[text width=15cm, text justified] at (1.5,-4.3){
{\hypertarget{Fig:1} \bf Fig.\,1}:
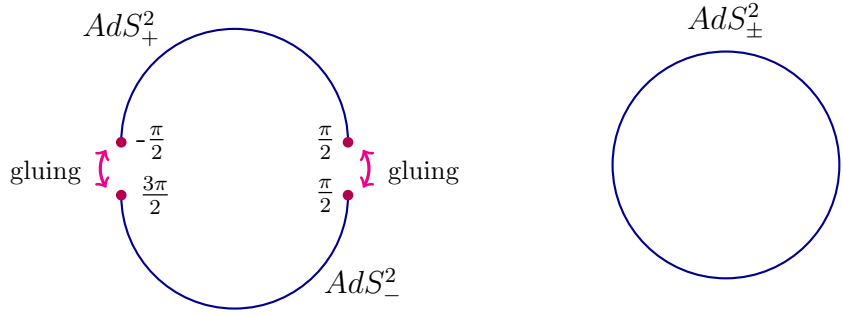
Construction of extended $AdS^2$ at fixed time.  
On the left, the two copies are glued along the boundaries. Each of these copies, denoted by $AdS^2_+$ and $AdS^2_-$, are coordinatized by $\theta$ which according to~\eqref{pmdef} runs over an interval of length $\pi$. 
On the right, the resulting space $AdS^2_\pm$ has no boundary and the fixed time topology  of a circle, with the coordinate $\theta$ running over an interval of length $2\pi$.};
\end{tikzpicture}
\end{center}

\subsection{Extended $AdS^3$}
\label{Sec22}
For the three-dimensional extension, we solve the hypersurface equation~\eqref{X2} by parametrizing the embedding space coordinates $(T^1, T^2, X^1, X^2)\in\R^{2,2}$ as
\begin{align}\label{eads3}
T^{1} &= L\sec\theta \cos(t/L)~,\qquad   
T^{2}= L\sec\theta \sin(t/L)~,  \cr
X^1  &= L \tan\theta \cos\phi~,~\quad\qquad
X^2  = L \tan\theta \sin\phi~,
\end{align}
where
\begin{equation}
0\leq\theta\leq\frac{\pi}{2}~, \qquad 0\leq\phi\leq2\pi, \qquad-\infty<t<\infty~.
\end{equation}
Note that in this case the polar coordinate $\theta$ ranges over an interval of length $\pi/2$, in contrast to the two-dimensional case where it spans an interval of length $\pi$. 

From~\eqref{gads} and~\eqref{eads3}, it follows that the induced AdS geometry is
\begin{equation}\label{gads3}
g_{AdS} = \frac{L^2 (\ext \theta^2+\sin^2\theta\,\ext\phi^2) - \ext t^2}{\cos^2\theta},
\end{equation} 
which in turn induces the topology of $D^2\times\R$, where $D^2$ is a two dimensional disk with coordinates $(\theta, \phi)$. The conformal boundary of $AdS^3$ is located at the boundary of disk~$\theta=\frac{\pi}{2}$ and has the topology of $S^1\times\mathbb R$.

The metric~\eqref{gads3} is invariant under the shift $\theta\to\pi-\theta$. This symmetry amounts to taking 
\begin{align}\label{pm3}
AdS^3_+&:=\bigg\{ (T^1, T^2, X^1, X^2)\in\R^{2,2} |~~0 \leq \theta \leq \frac{\pi}{2} \bigg\}~, \cr
AdS^3_-&:=\bigg\{(T^1, T^2, X^1, X^2)\in\R^{2,2} |~~\frac{\pi}{2} \leq \theta \leq \pi \bigg\}~,
\end{align}
as two different copies of $AdS^3$. These can be glued together along the boundary simply extending the range of the polar coordinate:
\begin{equation}
AdS^3_\pm = AdS^3_+ \cup AdS^3_- = 
\bigg\{(T^1, T^2, X^1, X^2)\in\R^{2,2} |~~0 \leq \theta \leq \pi \bigg\}~.
\end{equation}

It is important to point out that, strictly speaking, when gluing two manifolds $M$ and $N$ along their boundaries, one needs to specify the way the gluing is done via an invertible map $\varphi: \partial M \rightarrow \partial N$ such that the resulting space $M\cup_\varphi N$ is uniquely defined~\cite{LeeTopManifolds}. In our case, since we are considering two copies of the same manifold, it suffices to take $\varphi$ as the identity map; for simplicity we omit decorations regarding this point.

\subsection{Extended $AdS^{d+2}$}
We now formulate the general case of extended AdS space in dimension $d+2$, with $d\geq2$.  As we will see, the construction is a recursion that uses as a seed the lower-dimensional cases introduced above. 
To this end, we first parametrize
\begin{align}\label{eads}
T^1 &= L\sec\theta\,t^1,\qquad 
T^2 = L \sec\theta\, t^2,\cr
X^i &= L\sec\theta\, x^i  \quad  (i= 1,..., d-1), \cr
X^d &= L \tan\theta \cos\phi ,\qquad 
X^{d+1} = L \tan\theta \sin\phi~.  
\end{align}
where
\begin{equation}
0\leq\theta\leq\frac{\pi}{2}~, \qquad 0\leq\phi\leq2\pi~,
\end{equation}
and where $-\infty<(t^1, t^2, x^i)<\infty$ are coordinates in the embedding subspace $\R^{2, d-1}\subset\R^{2, d+1}$. These are given by the restriction 
\begin{equation}
(t^1, t^2, x^i) =  \frac{1}{L}  (T^1, T^2, X^i, X^{d}, X^{d+1})\Big\vert_{\theta=0}~.
\end{equation}

Next, we observe that the parametrization~\eqref{eads} solves the hypersurface equation~\eqref{X2} if and only if the coordinates $(t^1, t^2, x^i)$ satisfy the constraint
\begin{equation}\label{subX2}
-(t^1)^2 - (t^2)^2 + \sum_{i=1}^{d-1} (x^i)^2 =  - 1~.
\end{equation}
This constraint, which under~\eqref{eads} is the $\theta=0$ restriction of~\eqref{X2}, defines an hypersurface of negative curvature of codimension three with respect to the embedding space $\R^{2, d+1}$, and a submanifold of codimension embedded in $AdS^{d+2}$; we denote this submanifold by $\Sigma^d$.

To elucidate the significance of $\Sigma^d$ it is useful to compute the induced metric on the $AdS^{d+2}$ hyperboloid~\eqref{X2}. Making use of the parametrization~\eqref{eads} and imposing~\eqref{subX2}, a direct calculation yields 
\begin{equation}\label{ggen}
g_{AdS} = \frac{L^2(\ext\theta^2 + \sin^2\theta \,\ext\phi^2) + h}{\cos^2\theta}~,
\end{equation}
where 
\begin{equation}
h
=-(\ext t^1)^2 - (\ext t^2)^2 + \sum_{i=1}^{d-1} (\ext x^i)^2
\end{equation}
is the induced metric on $\Sigma^d$. Note that the explicit form of $h$ depends on the parametrization of the coordinates $(t^1, t^2, x^i)$, but for our purposes we do not need to make any particular choice of it. Global Einstein geometries of this type were first studied in~\cite{Arias:2019pzy, Arias:2023azh}.

From~\eqref{ggen} it follows that the induced topology on a single copy of AdS is 
\begin{equation}\label{top}
AdS^{d+2} \cong D^2 \times \Sigma^d~, 
\end{equation}
where $D^2$ is the two-dimensional disk with coordinates $(\theta, \phi)$, and $\Sigma^d$ is the submanifold defined by equation~\eqref{subX2}. Requiring $AdS^{d+2}$ to have a single, connected boundary we take $\Sigma^d$ to have no boundary so that 
\begin{equation}\label{top}
\partial (AdS^{d+2}) \cong S^1 \times \Sigma^d~.
\end{equation}
Since the constraint~\eqref{subX2} enforces $\Sigma^d$ to have negative curvature, we further take 
\begin{equation}\label{sigma}
\Sigma^d = AdS^d_\pm~,
\end{equation}
where we recall that the extended space $AdS^d_\pm$ has no boundary. 

The gluing of two copies of $AdS^{d+2}$ along their boundaries is almost identical to the three-dimensional case. Since we are dealing with a manifold of topology $D^2$ times a manifold without boundaries (the latter is given by $\R$ in the case of dimension 3, and by $AdS_\pm^d$ in the case of dimension $d+2$ ), then all we need to do is to glue together the two corresponding copies of $D^2$ along their $S^1$ boundary.  Indeed, due to the symmetry $\theta\to\pi-\theta$ exhibited by~\eqref{ggen} we can define 
\begin{align}\label{pm3}
AdS^{d+2}_+&:=\bigg\{ (T^1, T^2, X^i, X^d, X^{d+1})\in\R^{2,d+1}~|~~0 \leq \theta \leq \frac{\pi}{2} \bigg\}~, \cr
AdS^{d+2}_-&:=\bigg\{(T^1, T^2, X^i, X^d, X^{d+1})\in\R^{2,d+1}~|~~\frac{\pi}{2} \leq \theta \leq \pi \bigg\}~,
\end{align}
where both copies have the boundary located at $\theta=\pi/2$ (the point where the defining function $u=\cos\theta$ vanishes). Then, the extended space is directly obtained by extending the range of the polar coordinate as 
\begin{equation}\label{egen}
AdS^{d+2}_\pm = AdS^{d+2}_+ \cup AdS^{d+2}_- = 
\bigg\{(T^1, T^2, X^i, X^d, X^{d+1})\in\R^{2,d+1}~|~~0 \leq \theta \leq \pi \bigg\}~.
\end{equation}
Consequently and due to~\eqref{ggen}-\eqref{egen}, we observe that the glued, extended space can then be thought of as the recursion
\begin{equation}\label{rec}
AdS^{d+2}_\pm = S^2 \times AdS^d_\pm~, \quad d\geq2~,
\end{equation}
with seeds the lower dimensional cases $AdS^2_{\pm}$ and $AdS^3_\pm$ spelled out in \S\ref{Sec21} and \S\ref{Sec22}, respectively. A sketch of the construction is displayed in figure~\hyperlink{Fig:2}2. 
\vspace{7pt}
\begin{center}
\begin{tikzpicture}
\draw[thick, darkblue] (0,0) arc[start angle=0, end angle=180, radius=1.5cm];
\draw[thick, darkblue] (-3,-0.7) arc[start angle=180, end angle=360, radius=1.5cm];
\draw [thick, darkred](-1.5,0) ellipse (1.5cm and 0.2cm);
\draw [thick, darkred](-1.5,-0.7) ellipse (1.5cm and 0.2cm);
\node at (5,1.6) {$AdS^{d+2}_\pm$};
\draw[thick, darkblue] (5,-0.3) circle (1.5);
\draw [thick, darkred, dashed](5,-0.3) ellipse (1.5cm and 0.2cm);
\draw [magenta,<->, very thick] (-3.2,-0.1) to [bend right] (-3.2,-0.6);
\node at (-4,-0.4) {\footnotesize{gluing}};
\node at (-3,1.5) {$AdS^{d+2}_+$};
\node at (0.3,-1.9) {$AdS^{d+2}_-$};
\small
\node[text width=15cm, text justified] at (1.5,-3.7){
{\hypertarget{Fig:2} \bf Fig.\,2}: 
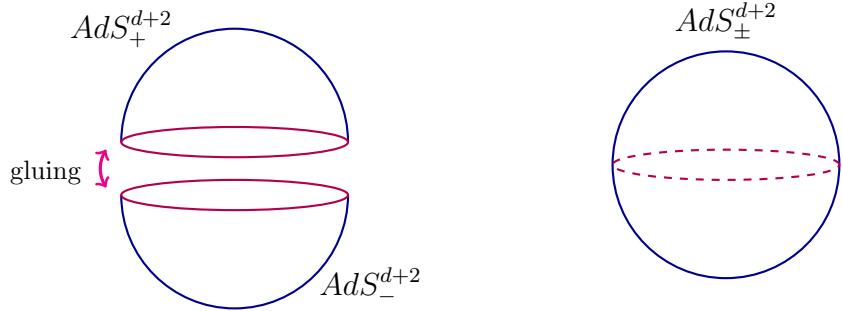
Gluing of two copies of $AdS^{d+2} \cong D^2 \times \Sigma^d$.  On the left, the corresponding disk factor of each copy glued along their $S^1$ boundary, located at $\theta=\pi/2$.  The resulting manifold has the topology of a 2-sphere, depicted on the right.
};
\end{tikzpicture}
\end{center}

\section{Coupling membranes}
\label{Sec3}
%
We now examine how the geometry of extended AdS space naturally admits the coupling of membranes.  
In what follows we consider $d\geq2$ and denote the metric on the extended space $AdS_\pm^{d+2}$  by $g_\pm$. This is given by~\eqref{ggen} but with an extended polar coordinate, that is
\begin{equation}\label{gpm}
g_\pm = \frac{L^2(\ext\theta^2 + \sin^2\theta \,\ext\phi^2) + h}{\cos^2\theta}~, \qquad
0\leq\theta\leq\pi~.
\end{equation}
It is important to recall that the line element above exhibits the shift symmetry $\theta\to\pi-\theta$ and induces the topology of $S^2\times \Sigma^d$ on $AdS^{d+2}_\pm$, where $\Sigma^d$ is a manifold of negative curvature without boundary that we take to be $AdS^{d}_\pm$; this choice yields the recursive definition~\eqref{rec} presented earlier.

\subsection{Antipodal defects}
The backreaction of codimension two cosmic strings and membranes generates conical defects~\cite{Vilenkin:1981zs}. These are characterized by a deficit angle which is in turn related to the tension of the string or membrane.

In our case, due to the induced topology of the extended space $AdS^{d+2}_\pm \cong S^2\times \Sigma^d$, one can straightforwardly induce a pair of conical defects by means of the quotient $S^2/\Z_n$; its fixed points are located at the north and south poles of the 2-sphere, and each of them support a singular surface that we will denote by $\Sigma^d_+$ and $\Sigma^d_-$, respectively.

Indeed, the action of $\Z_n$ can be defined, upon analytic continuation of the orbifold parameter $n\geq1$, by the azimuthal identification
\begin{equation}
\phi\sim \phi + \frac{2\pi}{n}~,
\end{equation} 
under which, locally about $\theta\to0, \pi$ and in terms of the variable $\rho=L\theta$, the metric~\eqref{gpm} behaves to leading order in $\rho$ as
\begin{equation}
g_\pm\approx \ext \rho^2 + \frac{\rho^2}{n^2} \ext \phi^2 + h~.
\end{equation} 
The latter manifestly exhibits the presence of conical defects at the north and south poles of the 2-sphere, both having a deficit angle given by $\Delta\phi=2\pi (1- \frac{1}{n})$. 
This maneuver amounts to thinking of the extended space $AdS^{d+2}_\pm$ as the smooth, $n\to1$ limit of the product $S^2/\Z_n\times \Sigma^d$. As anticipated, for $n>1$ we denote the two singular surfaces supported by the antipodal defects by
\begin{align}\label{sigmas}
\Sigma^d_+   =   (S^2/\Z_n\times \Sigma^d )\big\vert_{\theta=0} \qquad{\rm and}\qquad
\Sigma^d_-= (S^2/\Z_n\times \Sigma^d) \big\vert_{\theta=\pi}~.
\end{align}
Since the extrinsic curvatures of $\Sigma^d_+$ and $\Sigma^d_-$ vanish, these surfaces are minimal.

\subsection{Coupling an arbitrary number of membranes}
\label{Nmem}

Invariants of metrics on manifolds with conical defects, locally of the type  $\mathcal M=\mathcal C\times \Sigma$ where $\mathcal C$ is a cone,  have been studied in~\cite{Fursaev:1995ef}. The non-trivial holonomy about the conical singularity yields a Riemann tensor  containing a delta function with support on the codimension two singular surface $\Sigma$. This translates into a Ricci scalar which decomposes as
\begin{equation}\label{Rn}
 R^{(n)} = R - 4\pi \bigg(1-\frac{1}{n}\bigg) \delta_\Sigma 
\end{equation}
where $R=R^{(1)}$ is the curvature scalar of the smooth manifold and $\int_{\mathcal M} f\delta_\Sigma =\int_\Sigma f$. If the manifold $\mathcal M$ has globally a number of conical defects supporting several singular surfaces the last term in~\eqref{Rn} should be replaced by a sum over all such surfaces. 

Regarding the geometry of the extended space $AdS^{d+2}_\pm$, since we have the two singular surfaces~\eqref{sigmas} supported by antipodal conical defects of the same deficit angle, decomposition~\eqref{Rn} reads
\begin{equation}\label{Rnpm}
 R^{(n)} = R - 4\pi \bigg(1-\frac{1}{n}\bigg) (\delta_{\Sigma_+} + \delta_{\Sigma_-}). 
\end{equation}

The relation above has an immediate consequence when considering the gravitational action on the extended manifold with defects, which we recall has the induced topology of $S^2/\mathbb Z_n \times \Sigma^d$.  Indeed, when integrating~\eqref{Rnpm} one obtains the usual Einstein--Hilbert terms plus two extra Nambu--Goto terms with support on the singular surfaces~\eqref{sigmas}, that is 
\begin{align}\label{EH}
I[AdS_\pm/\mathbb Z_n] = \frac{1}{16\pi G} \int_{AdS_\pm} \ext^{d+2}x \sqrt{|{\rm det}\,g_\pm|}(R-2\Lambda)
&- \frac{1}{4G}\bigg(1-\frac{1}{n}\bigg)\int_{\Sigma_+} \ext^{d}y \sqrt{|{\rm det}\,h|} \cr
&-  \frac{1}{4G}\bigg(1-\frac{1}{n}\bigg)\int_{\Sigma_-}\ext^{d}y \sqrt{|{\rm det}\,h|}~,
\end{align}
where $y$ denotes the coordinates on $\Sigma_+$ and $\Sigma_-$.
Note that since the extended manifold $AdS^{d+2}_\pm$ is the result of gluing two copies of AdS space along their boundaries, the corresponding boundary terms in~\eqref{EH} cancel each other due to their opposite orientations.  Also note that, as Riemannian submanifolds, $\Sigma_+$ and $\Sigma_-$ are both endowed with the same induced metric $h$, and the last two terms in~\eqref{EH} are simply the area functional of the singular surfaces coupled through the tension 
\begin{equation}
\tau =\frac{1}{4G}\bigg(1-\frac{1}{n}\bigg)~.
\end{equation} 
The tensionless limit $n\to1$ corresponds to a smooth geometry without defects. 

It is crucial for our purposes to point out that, due to the isometries of extended AdS space, one can further couple an arbitrary number of coincident membranes to the action~\eqref{EH}. Indeed, instead of considering only the two singular surfaces~\eqref{sigmas}, one can extend the domain of the 2-sphere polar coordinate from $0\leq\theta\leq\pi$ to the interval
\begin{equation}\label{interval}
0\leq\theta\leq (2N-1)\pi~, \quad N\geq1~.
\end{equation}
As a result, the geometry $(AdS^{d+2}_\pm, g_\pm)$ exhibits $N$ coincident defects at the north pole of the 2-sphere, and $N$ coincident defects at the south pole, all of them supporting a total of $2N$ singular membranes; for finite $N$, the action~\eqref{EH} becomes
\begin{align}\label{EHN}
I[AdS_\pm/\mathbb Z_n] = \frac{1}{16\pi G} \int_{AdS_\pm} \ext^{d+2}x \sqrt{|{\rm det}\,g_\pm|}(R-2\Lambda)
&- \frac{N}{4G}\bigg(1-\frac{1}{n}\bigg) \Big[A(\Sigma_+) + A(\Sigma_-)\Big]~,
\end{align}
where $A(S)$ denotes the area of the hypersurface $S$. 
Evidently, $\theta$ can be extended as to run over the entire semi-positive real line $\theta\in [0, \infty]$, which translates into an infinite number of coupled membranes;  this limit will be essential in regularizing the entropy divergence in even dimensions.  A cartoon of the geometry with membranes is given in figure~\hyperlink{Fig:3}3.  
\vspace{20pt}
\begin{center}
\begin{tikzpicture}[scale=0.9]
\coordinate (N) at (-3.5,2);\coordinate (S) at (-3.5,-2);
\draw[thick] (N)to[out=-20,in=20](S);
\draw[thick] (N)to[out=-150,in=150](S);
\draw [thick] (-6.5,-2.2) --(-5.5,-1.7);
\draw [thick] (-6.5,-2.2) --(-1.5,-2.2);
\draw [thick] (-6.5,-2.4) --(-1.5,-2.4);
\draw [thick] (-6.5,-2.6) --(-1.5,-2.6);
\draw [thick] (-1.5,-2.2) --(-0.5,-1.7);
\draw [thick] (-1.5,-2.4) --(-0.5,-1.9);
\draw [thick] (-1.5,-2.6) --(-0.5,-2.1);
\draw [thick] (-5.5,-1.7) --(-0.5,-1.7);
\node [darkred] at (-3.5,-2) {\textbullet};
\draw [thick] (-6.5,2.2) --(-5.5,2.7);
\draw [thick] (-1.5,2.2) --(-0.5,2.7);
\draw [thick] (-5.5,2.7) --(-0.5,2.7);
\draw [thick] (-6.5,1.8) --(-1.5,1.8);
\draw [thick] (-1.5,2.0) --(-0.5,2.5);
\draw [thick] (-6.5,2.0) --(-1.5,2.0);
\draw [thick] (-6.5,2.2) --(-1.5,2.2);
\draw [thick] (-1.5,1.8) --(-0.5,2.3);
\node [darkred] at (-3.5,2) {\textbullet};
\node at (-5,1) {$AdS_+$};
\node at (-5,-1) {$AdS_-$};
\node at (-1.5,0) {$S^2/\mathbb Z_n$};
\draw[thick, dashed] (-2.4,0) 
arc[start angle=0,end angle=180, 
x radius=1.05, y radius=0.3];
\draw[thick] (-2.4,0) 
arc[start angle=0,end angle=-180, 
x radius=1.05, y radius=0.3];
%
\draw[thick, darkred, dashed] (5,2)to[out=-165,in=165](5,-2);
\draw[thick, darkred] (5,2)to[out=-15,in=15](5,-2);
\node [darkred] at (6.5,0) {$C$};
\draw [thick] (5,0) circle (2cm);
\node [darkred] at (5,2) {\textbullet};
\node [darkred] at (5,-2) {\textbullet};
\draw [thick] (2,2.2) --(3,2.7);
\draw [thick] (7,2.2) --(8,2.7);
\draw [thick] (3,2.7) --(8,2.7);
\draw [thick] (2,1.8) --(7,1.8);
\draw [thick] (7,2.0) --(8,2.5);
\draw [thick] (2,2.0) --(7,2.0);
\draw [thick] (2,2.2) --(7,2.2);
\draw [thick] (7,1.8) --(8,2.3);
\draw [thick] (2,-2.2) --(3,-1.7);
\draw [thick] (2,-2.2) --(7,-2.2);
\draw [thick] (2,-2.4) --(7,-2.4);
\draw [thick] (2,-2.6) --(7,-2.6);
\draw [thick] (7,-2.2) --(8,-1.7);
\draw [thick] (7,-2.4) --(8,-1.9);
\draw [thick] (7,-2.6) --(8,-2.1);
\draw [thick] (3,-1.7) --(8,-1.7);
\node at (-7.1,2) {$N\, \Big\{$};
\node at (-7.1,-2.4) {$N\, \Big\{$};
\node at (8.3,0.3) {$N\to\infty$};
\node at (8.3,-0.3) {$n\to1$};
\node[text width=16cm, text justified] at (0,-5) 
{\small {\hypertarget{Fig:3}\bf Fig.~3}: 
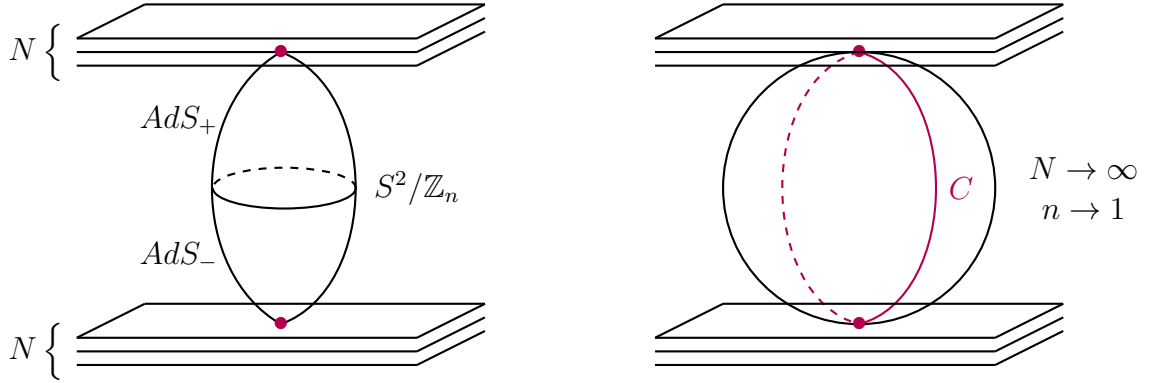
Extended AdS space without boundaries and with $2N$ coupled membranes. To the left, the two fixed points of the $S^2/\mathbb Z_n$ quotient support a stack of $N$ tensionful, coincident membranes, with the polar coordinates defined on the interval~\eqref{interval}.  To the right, the tensionless limit $n\to1$ of an infinite number of membranes on a smooth geometry.
};
\end{tikzpicture}
\end{center}

\section{Entanglement entropy}
\label{Sec4}
The coupling of an arbitrary number of membranes to the gravity action~\eqref{EHN} on the glued, extended space $AdS_\pm$ it would not have been possible on a single copy of AdS. 
The appearance of a new parameter, $N$, counting the number of membranes, amounts to taking the tensionless limit of these without neglecting their contribution to the action of the system. Indeed, it is direct to see that the Nambu--Goto terms in~\eqref{EHN} remain present in the tensionless limit $n\to1$ when $N\to\infty$, and vanish for finite $N$.

As we will next see,  the limit of a large number of membranes will play an important role when computing the entanglement entropy for an entangling surface consisting of the two antipodal stacks of membranes.
To begin with, we assume there exists some UV-complete gravity theory on the extended space $(AdS_\pm, g_\pm)$, whose low-energy limit is well-approximated by Einstein theory.
The classical approximation of the partition function of this theory is then given by the Euclidean gravity action
\begin{equation}\label{IR}
\log Z[AdS_\pm] \approx - I[AdS_\pm]~.
\end{equation}
Note that under this assumption and in this approximation, no details regarding the full quantum theory are required\footnote{While the full quantum dynamics of gravity on manifolds without boundary remains largely unexplored, this assumption is common in studies of gravitational entropy and is consistent with the effective field theory approach to quantum gravity, where one expects universal IR behavior irrespective of UV completion~\cite{Donoghue:1994dn}. The assumption has also proven to be useful in extracting reliable physical predictions at energy scales far below the Planck scale.}. 

We next consider a meridian circle $C$ at constant azimuthal angle, illustrated on the right of figure~\hyperlink{Fig:3}3, and take the bipartition $C=C_A\cup C_B$, where $C_A$ and $C_B$ are the two semicircles with boundary points the north and south poles of the original 2-sphere. Consequently, the product
\begin{equation}\label{Cauchy}
C\times \Sigma = (C_A\times\Sigma) \cup (C_B\times \Sigma)
\end{equation}
defines a bipartition into two regions that we assume to be entangled,  say $A=C_A\times\Sigma$ and $B=C_B\times \Sigma$, where the entanglement surface coincides with the two antipodal sets of membranes, that is 
\begin{equation}
\partial A=\partial B= \Sigma_+ \cup \Sigma_-~. 
\end{equation}
A similar entangling region, with an entangling surface given by two antipodal points on a holographic boundary 2-sphere, was considered in \cite{Donnelly:2018bef}.

We further attach to $C\times\Sigma$ a Hilbert space $H$ which factorizes\footnote{The assumption that the Hilbert space factorizes is known to be subtle in gauge theories and gravity, where local degrees of freedom cannot always be sharply localized. However, in semiclassical settings and for entanglement across geometrically separated regions, this assumption is often made as an effective approximation~\cite{VanRaamsdonk:2009ar, Donnelly:2014gva, Harlow:2016vwg}.} as $H= H_A \otimes H_B$, and consider a state $|\Psi\rangle \in H$ with reduced density matrix $\rho= \Tr_{H_B} | \Psi \rangle \langle\Psi|$ which we take to be normalized.

We are interested in computing the $n\to1$ limit of the Rényi entropies  
\begin{equation}\label{Ren}
S^{(n)}= \frac{1}{1-n} \log\T \rho^n~.
\end{equation}
To this end we employ the replica method~\cite{Calabrese:2004eu}, which in general gives  
\begin{equation}\label{trick}
\T \rho^n = \frac{Z[B_n]}{Z[B]^n}~,
\end{equation}
where $B_n$ denotes the $n$-fold branched cover of~\eqref{Cauchy}. In our case, since $C$ is a circle partitioned into two semicircles, the branched cover of $C\times \Sigma$ is built simply by gluing $n$ copies of $C\cong S^1$ along their north and south poles.  Crucially, the resulting manifold has the topology of $S^2\times \Sigma$, which is exactly the topology of the AdS extension~\eqref{rec}, that is 
\begin{equation}\label{voila}
B_n = AdS_{\pm} ~.
\end{equation}
Note that the branched cover $B_n$ is not singular but it admits the action of $\mathbb Z_n$, just as $AdS_\pm$ does. Of course, once this action is implemented the resulting manifold $B_n/\mathbb Z_n=AdS_\pm/\mathbb Z_n$ is conically singular.

From~\eqref{trick} and~\eqref{voila}, it follows that the entropies~\eqref{Ren} are given, in the semiclassical approximation~\eqref{IR}, by
\begin{align}\label{RenI}
S^{(n)}= \frac{1}{n-1} \Big( I[B_n] - n I[B]\Big)
= \frac{n}{n-1} \Big( I[AdS_\pm/\mathbb Z_n] - I[AdS_\pm]\Big)~,
\end{align}
where the last equality made use of the relation $I[B_n]=nI[B_n/\mathbb Z_n]$; it follows that the entanglement entropy 
\begin{equation}\label{SE}
S_{\rm E} = \lim_{n\to1} S^{(n)}  = n^2\partial_n  I[AdS_\pm/\mathbb Z_n]\Big\vert_{n=1}~.
\end{equation}

Consequently, the calculation of the entropy~\eqref{SE} reduces to the calculation of the on-shell, Euclidean action $ I[AdS_\pm/\mathbb Z_n]$ on the singular manifold $AdS_\pm/\mathbb Z_n$. As spelled out in~\S\ref{Nmem} and by means of the extension~\eqref{interval}, this action includes the coupling of $2N$ membranes.  Using the metric~\eqref{gpm} and the values 
\begin{equation}
R=-\frac{(d+1)(d+2)}{L^2}~,\quad
\Lambda=-\frac{d(d+1)}{2L^2}~,
\end{equation}
for the Ricci scalar $R$ and cosmological constant $\Lambda$, both in dimension $d+2$, the Euclidean\footnote{The Euclidean action carries an overall minus sign compared to the Lorentzian functional~\eqref{EHN}.} on-shell value of~\eqref{EHN} is
\begin{equation}\label{onshell}
 I[AdS_\pm/\mathbb Z_n]\approx
 \frac{d+1}{4nG} \bigg[\int_{0}^{\pi} \frac{\sin\theta}{|\cos\theta|^{d+2}} \ext\theta\bigg]  A(\Sigma)  
 +\frac{N}{4G}\bigg(1-\frac{1}{n}\bigg) \Big[A(\Sigma_+) + A(\Sigma_-)\Big]~.
\end{equation}
In the previous expression we have denoted by $A(\Sigma)=\int\ext y^d \sqrt{|{\rm det}\,h|}$, which, due to the form of the metric~\eqref{gpm} equals the areas $A(\Sigma_+)$ and $A(\Sigma_-)$.

Note that the integral in $\theta$ in the first term of~\eqref{onshell} blows up along the overlapped boundaries located at $\theta=\pi/2$; in terms of the defining function $u=\cos\theta$, the (principal value of the) singular integral is given by
\begin{equation}\label{singint}
\int_{0}^{\pi} \frac{\sin\theta}{|\cos\theta|^{d+2}} \ext\theta 
= \int_{-1}^{-\varepsilon} \frac{\ext u}{|u|^{d+2}} 
+ \int_{\varepsilon}^{1} \frac{\ext u}{|u|^{d+2}} 
= \frac{2}{d+1}\bigg(\frac{1}{\varepsilon^{d+1}}-1\bigg) 
\end{equation}
where $\varepsilon>0$ is a boundary cutoff. 

By virtue of the integral~\eqref{singint}, the on-shell action~\eqref{onshell} becomes
\begin{equation}
I[AdS_\pm/\mathbb Z_n]\approx
\bigg[\frac{1}{n}\bigg(\frac{1}{\varepsilon^{d+1}}-1\bigg) 
+N\bigg(1-\frac{1}{n}\bigg) \bigg]\,
\frac{A(\Sigma_+) + A(\Sigma_-)}{4G}~.
\end{equation}
It follows that the entropy~\eqref{SE} in this case is given by
\begin{equation}
S_{\rm E} = \bigg(1+N-\frac{1}{\varepsilon^{d+1}}\bigg)   \frac{A(\Sigma_+ \cup \Sigma_-)}{4G}~.
\end{equation}

We observe that the entropy above depends on the boundary cutoff $\varepsilon$ in a way that resembles the usual leading, short-distance divergence of any entanglement entropy in dimension $d+2$, namely $1/\varepsilon^{d+1}$.  However, in the present case, the freedom to couple an arbitrary number of membranes to the system amounts to take the double limit $N\to\infty$ and $\varepsilon\to0$ in a controlled way such that the divergences $N-1/\varepsilon^{d+1}$ cancel each other.  To this end, it suffices to take $N=10^k$ and $\varepsilon=10^{-k/(d+1)}$, with $k\to\infty$.

It is important to point out that, according to~\eqref{interval}, the $N\to\infty$ limit requires to extend the interval of the polar coordinate $\theta$ to the whole semi-positive real line. It is also crucial to observe that this cancelation of the divergences is only consistent in the tensionless limit $n\to1$, in which the coupling constant $N(1-1/n)$ in the action~\eqref{EHN} remains finite; when this is the case the entanglement entropy  
\begin{equation}
S_{\rm E} =  \frac{A(\Sigma_+ \cup \Sigma_-)}{4G}~,
\end{equation}
which is exactly the area law for the set of fixed points $\Sigma_+ \cup \Sigma_-$.

Finally, we would like to emphasize that although the above mechanism for canceling divergences may appear fine-tuned, it reflects a physically motivated balance. In the tensionless limit $n\to1$, the membranes become non-dynamical, yet their collective backreaction survives in the large-$N$ limit. 
This suggests that the extended geometry continues to encode a nontrivial contribution to the entanglement entropy, while avoiding the short-distance singularities that typically arise in local quantum field theory.
\section{Conclusions}
\label{Sec5}
In this work, we introduced a novel geometric framework for studying entanglement entropy in gravitational systems that do not rely on holography. By gluing together two copies of pure AdS spacetime along their conformal boundaries we constructed a manifold without boundary, that we denoted by $AdS^{d+2}_\pm$, and that exhibits the topology of $S^2\times\Sigma^d$, where $\Sigma^d$ is a negatively curved manifold without boundary. 
This extended space admits a natural coupling to codimension-two membranes, which appear as fixed-point minimal surfaces under the action of a discrete rotational symmetry acting on the 2-sphere.

Within this setup, we studied the semiclassical entanglement entropy across an entangling surface defined by two antipodal stacks of $N$ coincident membranes. 
Our main result is that the entropy obeys a universal area law, analogous to the Ryu–Takayanagi formula, yet derived entirely from bulk gravitational dynamics, without invoking any boundary dual description. Although a UV divergence inevitably arises from the near-boundary region, we showed that this divergence can be canceled in a controlled manner by taking the limit of an infinite number of membranes, thereby regularizing the entropy and restoring the area law. This large-$N$ mechanism for canceling short-distance divergences is the main insight of this work.

This construction provides a demonstration that area laws can arise from the intrinsic geometry of the bulk spacetime alone, independent of the existence of black holes or a holographic correspondence. 
The possibility of capturing Ryu–Takayanagi-type entropy purely from within the gravitational path integral on a manifold without boundary suggests a more universal origin for the relation between entanglement entropy and minimal surfaces, one that goes beyond holography.

There are several directions that would be worth exploring further. 
In particular, it remains an open question whether the large-$N$ mechanism responsible for the cancellation of divergences admits a deeper interpretation---arguably in terms of coarse-graining over microscopic degrees of freedom localized at the defects.
Another important question is whether this setup can be embedded in, or derived from, a UV-complete theory of gravity. It would be interesting to investigate whether the extended AdS geometry introduced here can be realized, for instance, as the near-horizon limit of a brane configuration in string theory or M-theory.

Ultimately, the results presented here support the idea that entanglement entropy and area laws may be deeply rooted in the geometric and topological structure of spacetime, regardless of whether a boundary theory exists. Our doubled AdS construction offers a concrete and tractable model in which these ideas can be further tested and developed.

\section*{\large Acknowledgements} 
This work was supported by the grant {\sc Fondecyt Postdoctorado} N$^{\rm o}$ 3220236, hosted by PUC Chile. 
\small
\bibliographystyle{JHEP}
\bibliography{AdSRefs}

\providecommand{\href}[2]{#2}\begingroup\raggedright\begin{thebibliography}{10}

\bibitem{Bekenstein:1973ur}
J.~D. Bekenstein, \emph{{Black holes and entropy}},
  \href{https://doi.org/10.1103/PhysRevD.7.2333}{\emph{Phys. Rev.} {\bfseries
  D7} (1973) 2333--2346}.

\bibitem{Hawking:1974sw}
S.~W. Hawking, \emph{{Particle Creation by Black Holes}},
  \href{https://doi.org/10.1007/BF02345020, 10.1007/BF01608497}{\emph{Commun.
  Math. Phys.} {\bfseries 43} (1975) 199--220}.

\bibitem{Maldacena:1997re}
J.~M. Maldacena, \emph{{The Large N limit of superconformal field theories and
  supergravity}}, \href{https://doi.org/10.1023/A:1026654312961,
  10.4310/ATMP.1998.v2.n2.a1}{\emph{Int. J. Theor. Phys.} {\bfseries 38} (1999)
  1113--1133}, [\href{https://arxiv.org/abs/hep-th/9711200}{{\ttfamily
  hep-th/9711200}}].

\bibitem{Ryu:2006bv}
S.~Ryu and T.~Takayanagi, \emph{{Holographic derivation of entanglement entropy
  from AdS/CFT}},
  \href{https://doi.org/10.1103/PhysRevLett.96.181602}{\emph{Phys. Rev. Lett.}
  {\bfseries 96} (2006) 181602},
  [\href{https://arxiv.org/abs/hep-th/0603001}{{\ttfamily hep-th/0603001}}].

\bibitem{VanRaamsdonk:2010pw}
M.~Van~Raamsdonk, \emph{{Building up spacetime with quantum entanglement}},
  \href{https://doi.org/10.1142/S0218271810018529}{\emph{Gen. Rel. Grav.}
  {\bfseries 42} (2010) 2323--2329},
  [\href{https://arxiv.org/abs/1005.3035}{{\ttfamily 1005.3035}}].

\bibitem{Lewkowycz:2013nqa}
A.~Lewkowycz and J.~Maldacena, \emph{{Generalized gravitational entropy}},
  \href{https://doi.org/10.1007/JHEP08(2013)090}{\emph{JHEP} {\bfseries 08}
  (2013) 090}, [\href{https://arxiv.org/abs/1304.4926}{{\ttfamily 1304.4926}}].

\bibitem{Calabrese:2004eu}
P.~Calabrese and J.~L. Cardy, \emph{{Entanglement entropy and quantum field
  theory}}, \href{https://doi.org/10.1088/1742-5468/2004/06/P06002}{\emph{J.
  Stat. Mech.} {\bfseries 0406} (2004) P06002},
  [\href{https://arxiv.org/abs/hep-th/0405152}{{\ttfamily hep-th/0405152}}].

\bibitem{Bombelli:1986rw}
L.~Bombelli, R.~K. Koul, J.~Lee and R.~D. Sorkin, \emph{{A Quantum Source of
  Entropy for Black Holes}},
  \href{https://doi.org/10.1103/PhysRevD.34.373}{\emph{Phys. Rev. D} {\bfseries
  34} (1986) 373--383}.

\bibitem{Srednicki:1993im}
M.~Srednicki, \emph{{Entropy and area}},
  \href{https://doi.org/10.1103/PhysRevLett.71.666}{\emph{Phys. Rev. Lett.}
  {\bfseries 71} (1993) 666--669},
  [\href{https://arxiv.org/abs/hep-th/9303048}{{\ttfamily hep-th/9303048}}].

\bibitem{Casini:2009sr}
H.~Casini and M.~Huerta, \emph{{Entanglement entropy in free quantum field
  theory}}, \href{https://doi.org/10.1088/1751-8113/42/50/504007}{\emph{J.
  Phys. A} {\bfseries 42} (2009) 504007},
  [\href{https://arxiv.org/abs/0905.2562}{{\ttfamily 0905.2562}}].

\bibitem{Casini:2011kv}
H.~Casini, M.~Huerta and R.~C. Myers, \emph{{Towards a derivation of
  holographic entanglement entropy}},
  \href{https://doi.org/10.1007/JHEP05(2011)036}{\emph{JHEP} {\bfseries 05}
  (2011) 036}, [\href{https://arxiv.org/abs/1102.0440}{{\ttfamily 1102.0440}}].

\bibitem{Eisert:2008ur}
J.~Eisert, M.~Cramer and M.~B. Plenio, \emph{{Area laws for the entanglement
  entropy - a review}},
  \href{https://doi.org/10.1103/RevModPhys.82.277}{\emph{Rev. Mod. Phys.}
  {\bfseries 82} (2010) 277--306},
  [\href{https://arxiv.org/abs/0808.3773}{{\ttfamily 0808.3773}}].

\bibitem{Gong:2017crn}
Z.-X. Gong, M.~Foss-Feig, F.~G. S.~L. Brand{\~a}o and A.~V. Gorshkov,
  \emph{{Entanglement Area Laws for Long-Range Interacting Systems}},
  \href{https://doi.org/10.1103/PhysRevLett.119.050501}{\emph{Phys. Rev. Lett.}
  {\bfseries 119} (2017) 050501},
  [\href{https://arxiv.org/abs/1702.05368}{{\ttfamily 1702.05368}}].

\bibitem{LeeTopManifolds}
J.~M. Lee, \emph{Introduction to Topological Manifolds}.
\newblock Springer, 2nd~ed., 2011.

\bibitem{Arias:2019pzy}
C.~Arias, F.~Diaz and P.~Sundell, \emph{{De Sitter Space and Entanglement}},
  \href{https://doi.org/10.1088/1361-6382/ab5b78}{\emph{Class. Quant. Grav.}
  {\bfseries 37} (2020) 015009},
  [\href{https://arxiv.org/abs/1901.04554}{{\ttfamily 1901.04554}}].

\bibitem{Arias:2023azh}
C.~Arias, \emph{{Reconstructing the boundary of AdS from an infrared defect}},
  \href{https://doi.org/10.1103/PhysRevD.108.126005}{\emph{Phys. Rev. D}
  {\bfseries 108} (2023) 126005},
  [\href{https://arxiv.org/abs/2307.02771}{{\ttfamily 2307.02771}}].

\bibitem{Vilenkin:1981zs}
A.~Vilenkin, \emph{{Gravitational Field of Vacuum Domain Walls and Strings}},
  \href{https://doi.org/10.1103/PhysRevD.23.852}{\emph{Phys. Rev. D} {\bfseries
  23} (1981) 852--857}.

\bibitem{Fursaev:1995ef}
D.~V. Fursaev and S.~N. Solodukhin, \emph{{On the description of the Riemannian
  geometry in the presence of conical defects}},
  \href{https://doi.org/10.1103/PhysRevD.52.2133}{\emph{Phys. Rev.} {\bfseries
  D52} (1995) 2133--2143},
  [\href{https://arxiv.org/abs/hep-th/9501127}{{\ttfamily hep-th/9501127}}].

\bibitem{Donoghue:1994dn}
J.~F. Donoghue, \emph{{General relativity as an effective field theory: The
  leading quantum corrections}},
  \href{https://doi.org/10.1103/PhysRevD.50.3874}{\emph{Phys. Rev. D}
  {\bfseries 50} (1994) 3874--3888},
  [\href{https://arxiv.org/abs/gr-qc/9405057}{{\ttfamily gr-qc/9405057}}].

\bibitem{Donnelly:2018bef}
W.~Donnelly and V.~Shyam, \emph{{Entanglement entropy and $T \overline{T}$
  deformation}},
  \href{https://doi.org/10.1103/PhysRevLett.121.131602}{\emph{Phys. Rev. Lett.}
  {\bfseries 121} (2018) 131602},
  [\href{https://arxiv.org/abs/1806.07444}{{\ttfamily 1806.07444}}].

\bibitem{VanRaamsdonk:2009ar}
M.~Van~Raamsdonk, \emph{{Comments on quantum gravity and entanglement}},
  \href{https://arxiv.org/abs/0907.2939}{{\ttfamily 0907.2939}}.

\bibitem{Donnelly:2014gva}
W.~Donnelly, \emph{{Entanglement entropy and nonabelian gauge symmetry}},
  \href{https://doi.org/10.1088/0264-9381/31/21/214003}{\emph{Class. Quant.
  Grav.} {\bfseries 31} (2014) 214003},
  [\href{https://arxiv.org/abs/1406.7304}{{\ttfamily 1406.7304}}].

\bibitem{Harlow:2016vwg}
D.~Harlow, \emph{{The Ryu\textendash{}Takayanagi Formula from Quantum Error
  Correction}}, \href{https://doi.org/10.1007/s00220-017-2904-z}{\emph{Commun.
  Math. Phys.} {\bfseries 354} (2017) 865--912},
  [\href{https://arxiv.org/abs/1607.03901}{{\ttfamily 1607.03901}}].

\end{thebibliography}\endgroup
\end{document}